\documentclass[superscriptaddress,twocolumn,aps,prd,longbibliography]{revtex4-1}

\usepackage[caption=false]{subfig}
\usepackage{lipsum}  
\usepackage{mathtools}
\usepackage{lineno}
\usepackage{hyperref}
\usepackage{graphicx}
\usepackage[normalem]{ulem}
\graphicspath{{./figs/}}
\usepackage{amsmath,amssymb}
\hypersetup{
colorlinks = true,
urlcolor   = blue,
citecolor  = blue,
linkcolor = blue
}
\modulolinenumbers[5]

\usepackage{xcolor}

\renewcommand{\tensor}[1]{\mathrm{\mathbf{#1}}}
\newcommand{\av}[1]{\left\langle{#1}\right\rangle}
\newcommand{\avsmall}[1]{\langle{#1}\rangle}

\newcommand{\kdot}{\dot{k}}
\newcommand{\dvr}{\delta v_r}

\begin{document}

\preprint{APS/123-QED}

\title{The Lagrangian kinetic energy cascade in Rayleigh-B\'{e}nard convection}

\author{Robin Barta}
\affiliation{Institute of Aerodynamics and Flow Technology, German Aerospace Center (DLR), G\"{o}ttingen, Niedersachsen, Germany}
\affiliation{Technical University Ilmenau, Institute of Thermodynamics and Fluid Mechanics, Ilmenau, Germany}

\author{Claus Wagner}
\affiliation{Institute of Aerodynamics and Flow Technology, German Aerospace Center (DLR), G\"{o}ttingen, Niedersachsen, Germany}
\affiliation{Technical University Ilmenau, Institute of Thermodynamics and Fluid Mechanics, Ilmenau, Germany}

\author{Ron Shnapp}
\affiliation{Mechanical engineering department, Ben Gurion University of the Negev, Beer Sheva, Israel}

\newcommand{\Ra}{\mathrm{Ra}}
\renewcommand{\vec}[1]{\boldsymbol{#1}}


\begin{abstract}
Rayleigh-B\'{e}nard convection at high Rayleigh number exhibits turbulence superimposed on large-scale circulation. While buoyancy forces drive the flow at certain scales, how kinetic energy is transfers across the scales is not understood. Here, utilizing a Lagrangian description of the kinetic energy flux, we present experimental evidence of a split cascade where energy flows downwscale at small scales and upscale at large scales. The flow topology of these energy transfer events differ profoundly, and the transition between them occurs gradually, over a broad range of scales.
\end{abstract}

\pacs{pacs1}
\maketitle

\begin{figure*}
	\centering
	\includegraphics[width=\textwidth]{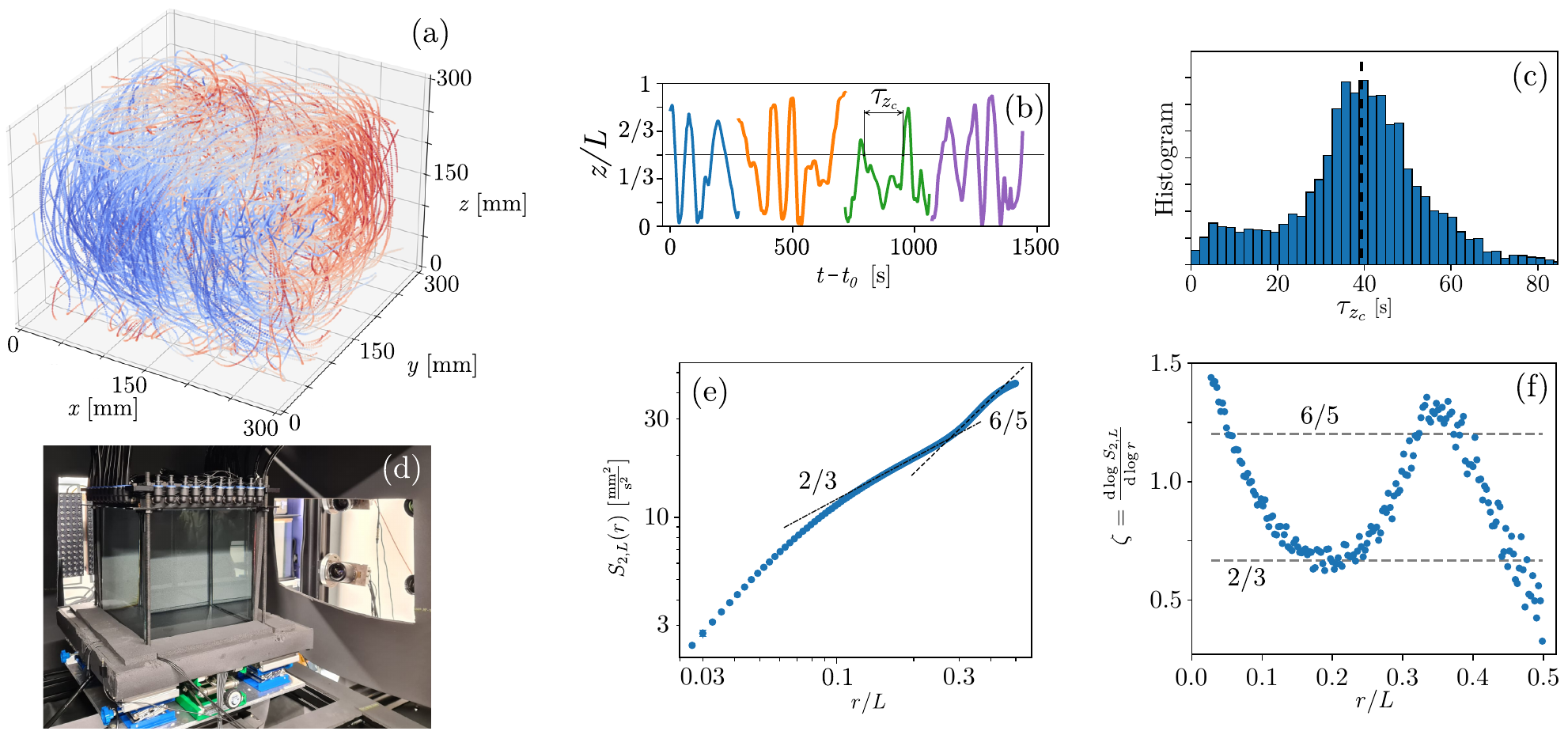}
	\caption{Characterization of the experimental system. (a) 3D rendering of approximately 2000 trajectories measured in the experiment. Colors represent the instantaneous vertical velocity where blue is directed with gravity and red is directed against it. (b) Examples of the $z$ coordinate of four tracer particles; the vertical line shows the center of the tank, $z_c=\frac{1}{2}L$ and $\tau_{z_c}$ is the time duration between two consecutive $z=z_c$ crossings. (c) Histogram of $\tau_{z_c}$ with a dashed line showing its mean value, $\av{\tau_{z_c}}=39\,\mathrm{s}$. (d) Picture of the experimental system including the water tank and, illumination and photography equipment used in the measurements. (e) Longitudinal second order structure function shown in log-log scales where its logarithmic slope is shown in (f). Two scaling law are shown corresponding to the K41 and BO59 theories.}
	\label{fig:system_fig}
\end{figure*}

\textit{Introduction--} 
Fluid density differences can drive fluid motion through conversion between potential and kinetic energies. This mechanism, so-called natural convection, is at the heart of numerous flows widely abundant in nature and industry.
One of the canonical examples is the Rayleigh-B\'{e}nard convection, in which heat supplied to the bottom of a fluid-filled vessel and extracted from its top lead to fluid flow. For sufficiently high Rayleigh numbers ($\Ra\gtrsim10^6$) turbulent flow is superimposed upon large-scale circulation (LSC)~\cite{Chilla2012, Pandey2018, LohseShishkina2024}. Due to the relative simplicity of its configuration and the complex physical phenomena it displays, the Rayleigh-B\'{e}nard convection has been widely used in the past to study the fundamental aspects of stratified turbulent flows (see~\cite{Lohse2024, LohseShishkina2024} and references therein).

A hallmark of turbulent flows is the energy cascade: kinetic energy is transferred from one scale to another through inertial flow structures~\cite{alexakis2018}. In the inertial range of three-dimensional turbulence the average energy flux is in the downscale direction---from larger to smaller scales. Thus, kinetic energy produced at large scales by an external forcing is transferred toward small scales to be dissipated there by viscous friction~\cite{Frisch1995}. In certain situations, such as in two-dimensional, rotational, and certain convective turbulent flows~\cite{kraichnan1967, Paret1997, Xia2009, Xia2011, Favier2014, Vieweg2022, Gallon2024}, the kinetic energy cascade can be reversed; there, the mean kinetic energy flux is in the upscale direction---from smaller to larger scales---and dissipation occurs due to friction of the large-scale flow structures against the domain boundaries.   
However, in turbulent Rayleigh-B\'{e}nard convection, the picture of the cascade is more complex~\cite{Lvov1991, Lohse2010}. Forcing is applied to the fluid through buoyancy forces of thermal plumes that detach from the hot and cold plates~\cite{Sparrow1970, kadanoff2001}, which can occur on a wide range of scales~\cite{alexakis2018}. Then, kinetic energy transfer across scales is split: it flows both downscale, and upscale~\cite{Togni2015, Vieweg2022}, dissipating due to viscous friction in the bulk and the boundary layers~\cite{Gayen2013}. 
Interestingly, a similar co-existence of LSC and turbulent fluctuations occur in buoyancy-driven bubbly flows~\cite{Shnapp2024}.
Overall, this split in energy pathways leads to the following questions: What are the characteristics of the transition (in scale) between downward and upward energy transfers, and how is this picture manifested in the dynamics of flow observables?

Energy fluxes in turbulence are often examined through the Eulerian approach, in which flow observables are analyzed in a reference frame fixed in space, e.g.~\cite{Haertel1994, eyink1995, Borue1998, Domaradzki2021, Park2025}. 
Here, however, we utilize the Lagrangian framework, analyzing flow observables along flow-tracing particles' trajectories. Advances in the Lagrangian description of turbulence have enhanced our understanding of transport and dispersion processes~\cite{shnapp2023, vieweg2024}, and energy transfer in turbulent flows~\cite{pumir2001, wan2010, meneveau2011, pumir2016}. In addition, conservation laws are often more naturally understood in the Lagrangian framework. Therefore, here we leverage the Lagrangian description to better understand turbulence dynamics in Rayleigh-B\'{e}nard convection. As we show below, there is a crossover in the direction of kinetic energy transfer across the scales from downscale at small scales to upscale at large scales. Detailed analysis shows that the transition is gradual, through an intermediate mixed regime, and that the flow structures associated with the downscale and upscale fluxes have different topology.\\

\textit{A Lagrangian view energy flux--} 
Temporal changes of tracer particles' kinetic energy are related to energy transfer in the flow~\cite{Mann1999, falkovich2001, pumir2001, xu2014, pumir2016, Gallon2024}. 
Under the Boussinesq approximation the acceleration of fluid particles is governed by the Navier-Stokes equation
\begin{equation}
\dot{\vec v} 
= \nabla\tensor\sigma +\vec f_b
= -\nabla p + 2\nu\,\nabla\tensor{S} + \vec f_b
\label{eq:NSE}
\end{equation}
where $\tensor\sigma\equiv -p\,\tensor I + 2\nu\,\tensor S$ is the stress tensor, $\vec f_b$ is the buoyancy force density, $\tensor S=\frac{1}{2}(\nabla \vec v + \nabla\vec v^{\top})$ is the rate of strain tensor, $p$ is the pressure divided by density and $\nu$ is the viscosity. Multiplying eq.~\eqref{eq:NSE} by $\vec v$ and averaging in space and time gives
\begin{equation}
\big\langle \frac{1}{2}\dot{\vec v^2} \big\rangle + \nabla\av{\vec T}  = \epsilon_b - \epsilon_k\,\, .
\label{eq:single_point_ke}
\end{equation}
Here, $\vec T \equiv p\,\vec v - 2\nu\,\vec v\, \tensor S $, and the average of its divergence is $\nabla\avsmall{\vec T}=0$ due to Stokes theorem.
The mean density of kinetic energy dissipation rate in eq.~\eqref{eq:single_point_ke} is
$\epsilon_k\equiv2\nu\,\av{\mathrm{Tr}(\tensor{S}\,\tensor{S})}$. In addition, mechanical work is applied to the fluid at a mean rate $\epsilon_b\equiv\av{\vec v\,\vec f_b}$. In statistically steady states $\avsmall{\dot{\vec v^2}}=0$, so the work and dissipation are equal $\epsilon_b = \epsilon_k$.

As our goal is to explore energy transfer across scales we consider the \textit{relative} motion between fluid tracers. 
The instantaneous relative velocity between two fluid tracers is $\delta_r \vec v \equiv \vec v(\vec x + \vec r, t) - \vec v(\vec x, t)$, and the kinetic energy of their relative motion is $k=\frac{1}{2}\delta_r \vec v^2$. 
Following Mann et al.~\cite{Mann1999}, we identify the mean rate of change of the relative motion kinetic energy, $\avsmall{\kdot}_r$, as the flux of kinetic energy from scales larger than $r$ to those smaller than $r$ (where $\av{\cdot}_r$ is a space and time average over all trajectory pairs with distance $\lvert \vec r \rvert = r$). 
Similarly, $\delta_r \vec v$ holds the minimum information needed to define the coarse-grained velocity and stress resolved at a scale $r$ which also supports identifying $\avsmall{\kdot}_r$ as the kinetic energy flux~\cite{Germano1992, pumir2001}.

To obtain an equation for $\avsmall{\kdot}_r$ we subtract eq.~\eqref{eq:NSE} at point $\vec x_1$ from itself at $\vec x_2$ with $|\vec x_1 - \vec x_2| = r$, and then multiply by $\delta_r \vec v$
\begin{equation}
\kdot =
(\vec v_1 - \vec v_2)\,(\nabla \sigma_1 - \nabla\sigma_2 +
\vec f_{b\,1} - \vec f_{b\,2})\,\, .
\label{eq:kdot_equation_pre_average}
\end{equation}
Averaging over time and space, and recalling that $\nabla\avsmall{\vec T}=0$, we obtain
\begin{equation}
\avsmall{\kdot}_r = 2\,\epsilon_{b} - 2\epsilon_{k} -
2\avsmall{\vec v \, \nabla\,\tensor \sigma^*}_r -
2\avsmall{\vec v \, \vec f_b^*}_r
\label{eq:kdot_equation}
\end{equation}
where terms of the form $\av{\vec v \, X^*}_r$ represent a covariance between the velocity at one point and a property $X$ at a point separated by distance $r$. 
As shown in eq.~\eqref{eq:single_point_ke}, the $\epsilon_k$ and $\epsilon_b$ terms cancel each other out. Furthermore, as the strain rate is a small scale property in turbulence~\cite{tsinober2009} we expect that $\avsmall{\vec v \, \nabla\,\tensor S^*}_r \approx 0$ for $r$ above the dissipation range ($r\gg\eta$ where $\eta\equiv(\nu^3/\epsilon_k)^{1/4}$ is the Kolmogorov scale). 
Overall, the flux results from spatial correlations between the fluid velocity and buoyancy force, and between the fluid velocity and pressure across distance $r$
\begin{equation}
\avsmall{\kdot}_r =
2\avsmall{\vec v \, \nabla\,p^*}_r -
2\avsmall{\vec v \, \vec f_b^*}_r
\qquad (\text{for}\quad \eta \ll r)  \,\,.
\label{eq:kdot_equation_remaimning_terms}
\end{equation}
In contrast, in the inertial range of homogeneous isotropic turbulence (HIT), $\avsmall{\dot{k}}_r = -2\epsilon_k$ for $\eta\ll r\ll L$ as the flux is constant~\cite{Mann1999,falkovich2001}.
%
%
\\

\begin{figure}[h]
	\centering
	\includegraphics[width=\linewidth]{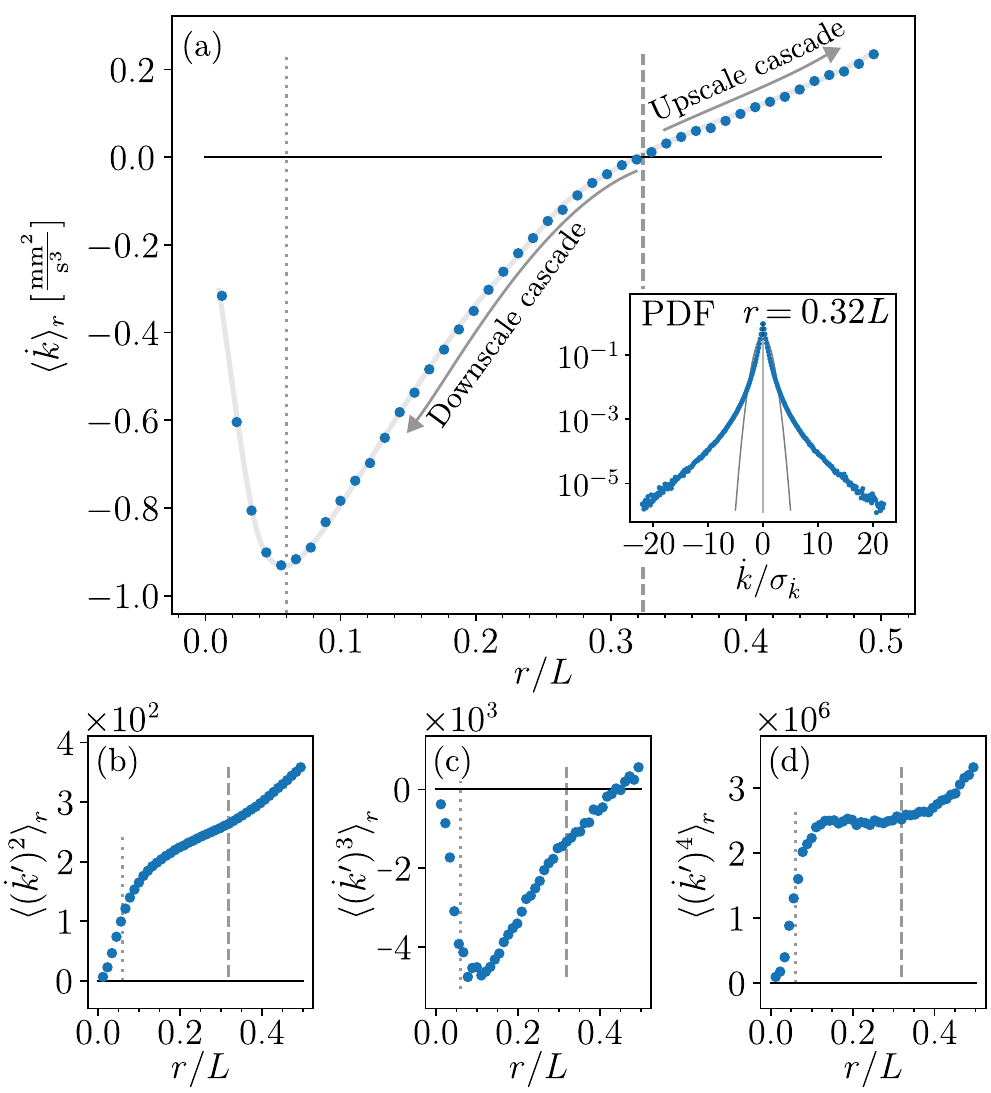}
	\caption{Statistics of $\dot{k}$ as a function of separation. Panel (a) shows the $r$ conditional mean of $\kdot$, and a PDF of $\kdot$ for $r=0.32L$ is shown in the inset and compared with a Gaussian distribution. Panels (b), (c) and (d) show the second, third and fourth central moments of $\kdot$ as a function of $r$ respectively. The dotted and dashed lines show the transitional values, $r^*$ and $r^\dagger$ respectively (see text).}
	\label{fig:lagrangian_cascade}
\end{figure}

\textit{Methods--} In what follows, we explore the statistics of the rate of change of the kinetic energy of the relative velocity ($\kdot$) in a Rayleigh-B\'{e}nard convection experimentally.
The flow in the experiment was in a cubic cell with side lengths $L=300$mm, containing water as the working fluid. The flow was heated from below and cooled from above with two aluminum plates kept at a constant temperature difference of $\delta T = 6.5^\circ\pm0.1^\circ\,\mathrm{C}$. Tracer particles with $20\,\mu\mathrm{m}$ mean diameter and density $1.03\,\mathrm{g}\,\mathrm{cm}^{-3}$ were seeded and tracked in the flow using proPTV--an open source 3D particle tracking velocimetry (3D-PTV) software~\cite{barta2024, Barta2025}.
In this work, we analyzed a dataset taken from a 30 hour experiment carried out to study the LSC~\cite{barta2024b}. The dataset we used for the analysis contained $\sim2.5\times10^5$ trajectories taken from time windows of approximately a 1.5 hours. Figure~\ref{fig:system_fig}a visualizes the flow with a 3D rendering of trajectories from the experiment. The Rayleigh number, characterizing the thermal forcing in the experiment was $\mathrm{Ra}=2.5\times10^9$, and the mean free-fall time, characterizing the average period it takes a Lagrangian tracer to complete one half of the circulation cycle, is approximately 40s (Fig.~\ref{fig:system_fig}b, c). Full experimental details are given in Ref.~\cite{barta2024b}.

To characterize the scales relevant for the turbulent flow in the experiment we show in Fig.~\ref{fig:system_fig}e the longitudinal velocity second order structure function,
\begin{equation}
S_{2,L}(r) = \av{\big(\delta_r \vec v \, \frac{\vec r}{r} \big)^2} \,\, ,
\end{equation}
and in Fig.~\ref{fig:system_fig}f we show its local logarithmic slope,
\begin{equation}
\zeta(r) = \frac{d\,\log \, S_{2,L}}{d\,\log \, r} \,\, .
\end{equation}
We obtained these results by binning samples of $\delta_r \vec v$ according to $r$, each bin holding several millions of samples (between $\sim3\times10^6$ for the smallest $r$ bin and up to $\sim50\times10^6$), and calculated statistics over these ensembles. During bin assembly we took samples from the whole region of measurement which consists the whole fluid tank excluding region very near to the walls. This technique was used to calculate the results discussed throughout this paper.
Mean field theory for Rayleigh-B\'enard convection predicts $S_{2,L}\propto r^{\zeta(r)}$, with three characteristic regimes~\cite{Lohse2010}. For $r\ll\eta$ the scaling $\zeta(r)=2$ corresponds to the dissipation range, which is not resolved in our experiment. For $ \eta \ll r \ll L_B$ an inertial range is expected with the K41 scaling of $\zeta(r)=\frac{2}{3}$ ($L_B$ is the Bolgiano scale). A narrow inertial range scaling is seen in our data for $0.15L \lesssim r \lesssim 0.25L$ in which $\zeta=0.67\pm0.06$. The lower end of this range suggests that the Kolmogorov scale in our flow is on the order of a millimeter $\eta\sim\mathcal{O}(1)\,\mathrm{mm}$, and the velocity dissipation rate is on the order of $\epsilon_k \sim \mathcal{O}(1)\, \mathrm{mm}^2\,\mathrm{s}^{-3}$. Above $L_B$ buoyancy forces become dominant and the Bolgiano-Obukhov theory (BO59)~\cite{Bolgiano1959} leads to $\zeta=\frac{6}{5}$. In Fig.~\ref{fig:system_fig}f, a narrow range of $\zeta=1.3\pm 0.07$ is seen for $0.3L \lesssim r \lesssim0.4L$, slightly higher than the BO59 scaling. The OB59 scaling is known to be elusive in empirical studies~\cite{kaczorowski2013}, and the higher value of $\zeta$ observed here can be attributed to shear effects associated with the LSC~\cite{Lohse2010}. All in all, we estimate the Bolgiano scale in our experiment to be approximately $L_B \approx 0.3L$.
\\

\textit{Flux statistics--} 
The statistical behavior of $\dot{k}$ as a function of $r$ in the Rayleigh-B\'{e}nard convection is characterized in Fig.~\ref{fig:lagrangian_cascade}. The distribution of $\dot k$ is non-Gaussian and has exponential tails (for example, Fig.~\ref{fig:lagrangian_cascade}a, inset). Consequently, extreme events of $\kdot$ being many standard deviations away from the mean occur at probabilities much higher relative to a normal distribution at all scales $r$ we tested. 
For example, the flatness factor of $\kdot$ ($F(r) = \avsmall{(\dot{k}')^4}_r \, / \, \avsmall{(\dot{k}')^2}_r^2$ where $\dot{k}' = \dot{k} - \avsmall{\dot{k}}_r$) decreased with $r$ from approximately 200 at $r=0.05L$ to approximately 30 at $r=0.5L$ (not shown); on the other hand, assuming the velocity and the acceleration were normaly distributed and uncorrelated we would expect the flatness factor of $\kdot$ to be 9 (as $\kdot=\sum_i v_i\,\dot{v}_i$). It is indeed a well-known feature of three dimensional turbulence that small scale properties, such as the acceleration, have a high flatness factor---a phenomenon called intermittency~\cite{arneodo2008, schumacher2009, shnapp2021}. Nevertheless, while in HIT the flatness of velocity increments decreases with $r$ throughout the inertial range and reaches close to Gaussian values ($F(50\eta)\approx4$ in HIT~\cite{tabeling1996}), the flatness of $\dot{k}$ observed here remained very high all the way up to $r=\frac{1}{2}L$, which is above the inertial range and above $L_B$ (Fig.~\ref{fig:system_fig}). We believe that the high flatness at such large scales is due to the signature of the driving mechanism in the Rayleigh-B\'enard cell, namely the thermal plumes.

We observe two transitions in the behavior of $\dot{k}$ as a function of $r$. The first transition, shown as dotted line in Fig.~\ref{fig:lagrangian_cascade}, is identified by the observation that for $r<r^*$ where $r^*\equiv (0.06\pm 0.013)L$ the mean $\avsmall{\dot{k}}_r$ is negative and it decreases with $r$, while for $r>r^*$, $\avsmall{\dot{k}}_r$ increases with $r$; the same is observed for the third moment, $\avsmall{(\dot{k}')^3}_r$. Furthermore, for $r<r^*$ the second and fourth moment of $\kdot$ increase sharply with $r$ while they grow more gradually with $r$ for $r>r^*$. 
In the inertial range of HIT we would expect the mean $\avsmall{\kdot}_r$ and the third moment $\avsmall{(\dot{k}')^3}_r$ to plateau with $r$ (as  $\avsmall{\dot{k}}_r = -2\epsilon_k$, and see~\cite{xu2014}), and such plateaus are not observed in our Rayleigh-B\'{e}nard flow (recall, the inertial range coincides with $0.15L\lesssim r \lesssim0.25L$, slightly above $r^*$). The growth of $\avsmall{\kdot}_r$ in the inertial range suggests that the downscale kinetic energy flux is not constant in the inertial range as it decreases with $r$.

The second transition, shown with dashed lines in Fig.~\ref{fig:lagrangian_cascade}, is associated with a crossover of $\avsmall{\kdot}_r$ from negative to positive values, at $r^\dagger=0.32\pm0.01\,L$. At the same scale, there is also an increase in the growth rate of the second and fourth moments with $r$ signaling on higher $|\kdot|$ values for $r>r^\dagger$, in addition to a slight decrease in the growth rate of the third moment of $\kdot$ with respect to $r$.  
Having a negative mean, $\avsmall{\kdot}_r<0$, is expected at small scales due to the downscale directed cascade~\cite{Mann1999,falkovich2001}. This transition roughly coincides with the Bolgiano scale, $L_B$. Thus, the transition to $\avsmall{\dot k}_r>0$ denotes a transition to upscale cascade for $r>L_B$ in which thermal plumes drive the LSC (eq.~\eqref{eq:kdot_equation}). This is in agreement with the split cascade observed in previous Eulerian studies~\cite{Togni2015, boffetta2012}. \\

\begin{figure}[h]
	\centering
	\includegraphics[width=\linewidth]{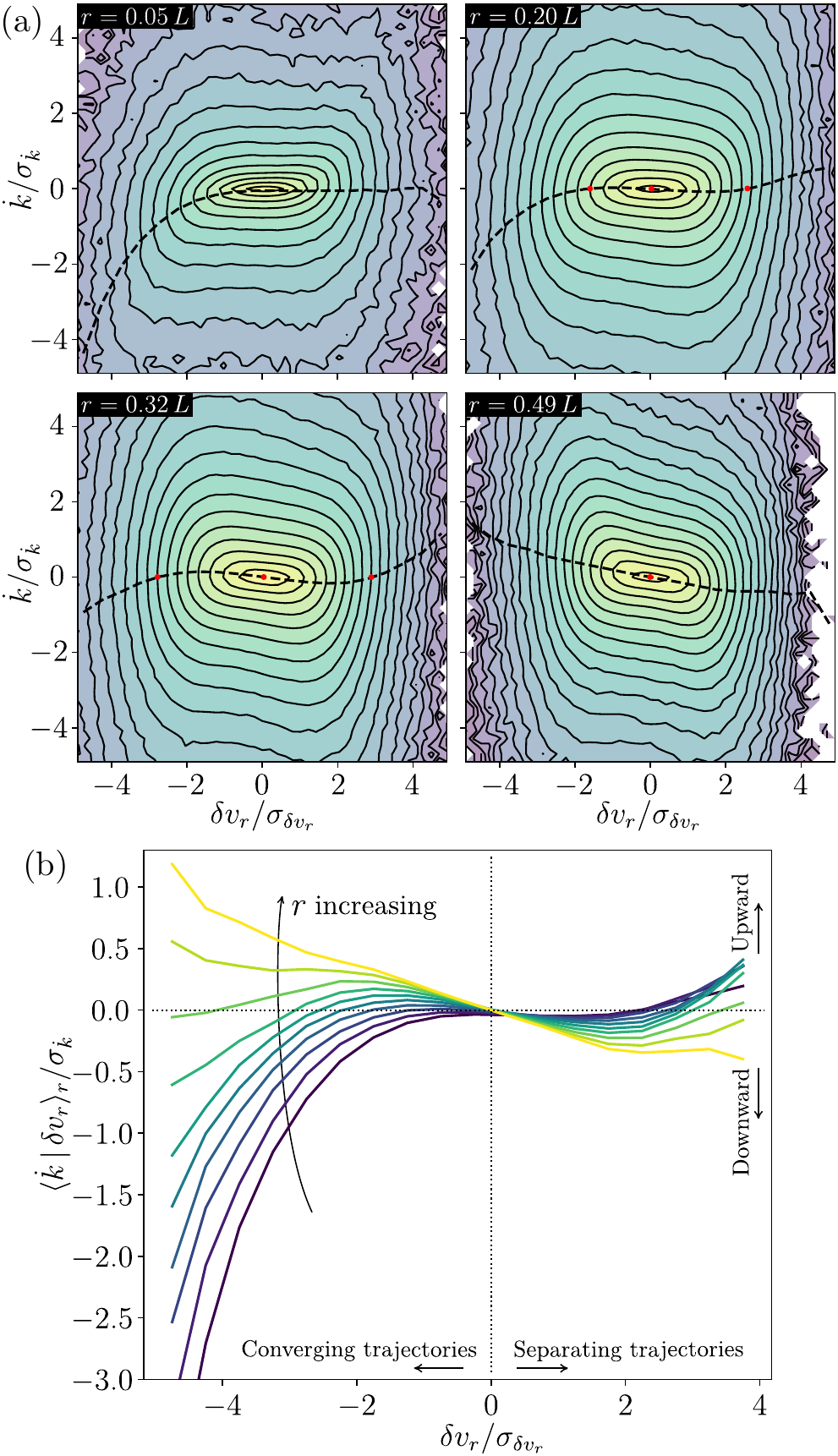}
	\caption{(a) Joint probability distribution functions of the relative kinetic energy rate of change and the particles' separation velocity normalized using their standard deviations. Data shown for four $r$ bins. Dashed lines show the conditional mean $\avsmall{\dot{k} \, \lvert \delta v_r}_r$ as a function of the separation velocity where red dots show its zero crossings. (b) Conditional mean $\avsmall{\dot{k} \, \lvert \delta v_r}_r$ for ten $r$ values uniformly spread starting from $r=0.078L$ and up to $r=0.47L$.}
	\label{fig:cascade_stretching}
\end{figure}

\textit{Flow structure and energy flux--}
The downward kinetic energy flux in three dimensional turbulence can occur through two processes: strain self-amplification and vortex stretching~\cite{Borue1998, tsinober2009}; these processes support downscale energy transfer if the strain skewness is negative and vortex stretching is positive respectively. Previous studies in three dimensional turbulence have shown that the downscale cascade is mostly correlated with straining regions~\cite{pumir2001, Borue1998} which suggests that strain self-amplification is the more dominant mechanism for downscale energy transfer. Furthermore, the flow structure around intense energy transfer events has a saddle-node topology where the coarse-grained rate-of-strain tensor has two positive and one negative eigenvalues, resulting in bi-axial strain~\cite{Park2025}. On the other hand, the flow topology responsible for the upscale transfer is much less known. Two mechanisms have been suggested: equal-sign vortex merger and vortex thinning by large scale strain, where the latter has received stronger support from numerical simulations~\cite{Chen2006, xiao2009}. Notably, these mechanisms were suggested for purely 2D turbulent flows which can be different from the split cascade in 3D flows~\cite{musacchio2017}.

To better understand the flow structure associated with the two transitions observed above we examine the joint distribution and conditional mean of $\kdot$ with the separation velocity between particles
\begin{equation}
\delta v_r \equiv \delta_r \vec v\,\frac{\vec r}{r} \,\,.
\end{equation}
Cases in which $\delta v_r<0$ correspond to converging trajectories while $\delta v_r>0$ means separating trajectories. Notably, in bi-axial strain there are two divergence axes versus only one convergence axis so in such a flow topology converging trajectories will tend to have higher $|\delta v_r|$ than diverging trajectories; in axial strain the opposite will occur and separating trajectories will tend to have higher $|\delta v_r|$. Thus, observing values of $\dvr$ can hint of topological aspects of the flow.

In Fig.~\ref{fig:cascade_stretching} we show joint statistics of $\kdot$ and $\delta v_r$ as a function of $r$. In the small scale regime, $r\lesssim r^*$, the joint PDF is biased towards negative $\kdot$ and $\dvr$ values. In particular, the conditional mean $\avsmall{\kdot \,\lvert \dvr}_r$ for $\dvr<0$ (namely for converging trajectories) is negative and it increases monotonically with $\dvr$, while it is approximately $\avsmall{\kdot \,\lvert \dvr}_r=0$ for $\dvr>0$ (namely separating trajectories). Because $\avsmall{\kdot}_r<0$ characterizes the downward cascade, this behavior supports the association of downscale kinetic energy flux with bi-axial flow topology for $r<r^*$. Namely, our results support the view that the Rayleigh B\'enard convection at small scales displays cascade characteristics similar to the three dimensional Kolmogorov turbulence.

For scales $r\gtrsim r^*$ and up to slightly above $r^\dagger$, there is a different behavior in which an inflection point appears for the conditional mean $\avsmall{\kdot \,\lvert \dvr}_r$ at $\dvr\approx0$ (Fig.\ref{fig:cascade_stretching}b). Strongly converging trajectories ($\dvr \ll 0$ events) are still associated with $\avsmall{\kdot \,\lvert \dvr}_r$ being negative, yet for weaker $\dvr<0$ trajectories $\avsmall{\kdot \,\lvert \dvr}_r$ is positive. For separating trajectories the opposite occurs: strongly separating trajectories ($\dvr\gg0$ events) are associated with $\avsmall{\kdot \,\lvert \dvr}_r$ being positive, while weaker separating trajectories are associated with $\avsmall{\kdot \,\lvert \dvr}_r$ being negative. These behaviors, characterized by the negative slope at $\dvr=0$, intensify as $r$ increases in this range.

Lastly, for $r\gg r^\dagger$ the inflection point disappears and $\avsmall{\kdot \,\lvert \dvr}_r$ decreases monotonically with $\dvr$ crossing zero from above at $\dvr\approx0$. Therefore, for scales $r \gg r^{\dagger}$, strongly converging trajectories are associated with $\avsmall{\kdot\lvert\dvr}_r>0$ while separating trajectories are associated with $\avsmall{\kdot\lvert\dvr}_r<0$. 
As $\avsmall{\kdot}_r$ characterizes the direction of kinetic energy flux,
this suggests that the flow structures associated with energy transfer at the large scales regime differ from those at the small-scale regime. Specifically, the mean energy transfer in this regime is in the upscale direction ($\avsmall{\kdot}_r>0$), and the events that contribute to this upscale transfer are more associated with converging trajectories. 
Furthermore, the downscale transfer events in this regime ($\kdot<0$) occur mostly in association with separating trajectories; this is in stark contrast with the observation above that downscale flux events at small scales are correlated with converging trajectories.\\

\begin{figure}
	\centering
	\includegraphics[width=\linewidth]{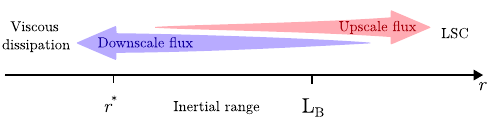}
	\caption{A conceptual sketch for the energy transfer regimes in Rayleigh-B\'{e}nard turbulence.}
	\label{fig:sketch}
\end{figure}

\textit{Discussion--} The transitions in the behavior of $\kdot$ with respect to scale thus correspond to three regimes of energy transfer in the Rayleigh-B\'enard flow. At small scales near the dissipation range the mean energy transfer is downscale, and, similar to HIT flows, the flux is associated with converging trajectories flow topology. At large scales, $r>L_B$, the energy transfer is mostly in the upscale direction and associated with converging trajectories, while the downscale energy transfer events in this regime are associated with separating trajectories. Therefore, in the small and large scale regimes both the mean direction of energy transfer and the correlation between flow topology and energy flux are reversed. This suggests that the flow structures associated with downscale and upscale energy transfer are not simply time-reversed images of one-another, so they occur due to different physical mechanisms. 
The third regime, $r^*<r<L_B$, is a mixed inertial regime laying between the former two; there, intense energy transfer events (namely farther from the mean) are of the type observed in the HIT-like downscale cascade, while the less intense events are of the type observed in upscale cascade regime. The intermediate regime manifests a smooth transition between the upscale-downscale cascades. A possible explanation for the behavior observed, presented conceptually in Fig.~\ref{fig:sketch}, is that downscale and upscale transfer events occur simultaneously over a wide range of inertial scales above $r^*$ with different proportions, and these balance each other out around $L_B$.
\\

\textit{Conclusions--} This work explores the transfer of kinetic energy across scales in Rayleigh-B\'enard convection through the Lagrangian framework. We observe three energy transfer regimes based on scale. At small scales the mean energy flux is in the downward direction while at large scale the mean flux is in the upward direction. The third regime is an intermediate one that mixes the behaviors of the former two where more extreme events were correlated with downscale transfer while less intense events were associated with upscale transfer. Furthermore, the flow topology observed for the different downscale and upscale cascade regimes was different in the small and large scales regimes. This shows that split cascade turbulent flows present a mixture of inertial energy transfer mechanisms that occur simultaneously but are separated in scale-space.
Furthermore, it shows that the physical mechanisms responsible for the down- and upscale energy transfer turbulent flows are not simply time-reversed images, but are profoundly different from one another, thus highlighting the irreversibly associated with turbulent flows.

\begin{acknowledgements}
	
\noindent The authors thank Prof. Alain Pumir for insightful comments on the initial draft. RS acknowledges support from ISF grants 1244/24 and 2586/24 and from the Alon Scholarship.
	
\end{acknowledgements}

\bibliography{bibliography}

\end{document}